\documentstyle[12pt]{article}
\begin{document}
\begin{center}
\noindent {\large {\bf Behaviour of Magnetic Tubes in Neutron Star's Interior}}\\
\vspace{0.5cm}
${R.S. Singh}^1$, ${B.K. Sinha}^2$ and ${N.K. Lohani}^3$\\
\vspace{0.5cm}
${1.}_{Department of Mathematics, P.G. College,Ghazipur (U.P.), India}$\\
${2.}_{Department of Mathematics, S.C. College,
Ballia (U.P.), India }$\\
${3.}_{Department of Physics, M.B. Govt. P.G. College, Haldwani, Nainital - 263 141, India}$\\
\end{center}
\vspace{0.2cm}
\noindent {\bf Abstract.} It is found from Maxwell's equations that the 
magnetic field lines are good analogues of relativistic strings. It is shown 
that the super-conducting current in the neutron star's interior causes local rotation of magnetic flux tubes carrying quantized flux.\\ 
\begin{center}
\section{\bf Introduction}
\end{center}
\noindent It is widely accepted that neutron star's interior is mostly neutrons
 with a small admixture of protons and an equal number of electrons [1]. The  electrons form a highly degenerate relativistic plasma which is in the normal 
 state [2]. The magnetic flux that threads the interior is generated by the 
 electron currents (i.e., the normal conduction currents). As the temperature 
 of neutron star's interior drops below the critical temperature ${\bf T}_c$, protons 
 form a type II super-conductor in which the magnetic field is organized into 
 an array of quantized magnetic flux tubes [3]. The super-conducting electron
 revolve around these tubes and generate superconducting currents. Since 
 protons are super-conducting, the diffusion time scale for the magnetic field 
 is lengthened and hence the field in the interior can be thought of as frozen-in  on a macroscopic scale. Lagrangian description of the frozen-in magnetic  field leads to the action associated with Nambu-Goto strings [4]. This result 
 gives us impetus to investigate the string behaviour of magnetic field lines. 
 In this short note we derive string equations of motion from Maxwell's equations and prove that the quantized magnetic flux tubes in the super-conducting region are locally rotating tubes.\\   

\section{\bf String Behaviour of Magnetic Field Lines}

\vspace{0.2cm}
\noindent We consider a highly conducting fluid with 4-velocity ${{\bf u}^a}$ which satisfies the normalization condition ${{\bf u}^a}$ ${{\bf u}_a}$ = -1. The electro-magnetic field   inside   the  fluid  evolves  in accordance with Maxwell's equations [5]. \\

\begin{equation}
{\bf H}_{[ab;c]} = 0 \hspace{1.0cm}  or \hspace{1.0cm} {{\bf H}^{*ab}_{;b}} = 0 ,
\end{equation}
\begin{equation}
{\bf H}^{ab}_{;b} = {{\bf J}^a},
\end{equation}
\noindent where semi-colon (;) indicates covariant derivative and asterisk (*) denotes dualization of 2-form. ${\bf J}^a$ is the electric current 4-vector which is expressible as \\
\begin{equation}
{\bf J}^a = {\bf q}{\bf u}^a + {\bf \jmath}^a_{(n)} + {\bf \jmath}^a_{(s)},
\end{equation}
\noindent where ${\bf q}$ is the proper electric charge density. ${\jmath}^a_{(n)}$ and ${\jmath}^a_{(s)}$  denote, respectively, the normal conduction current and the super-conducting current. In the limit of infinite electrical conductivity, the electric field ${\bf E}_a$ measured by an observer comoving with the fluid vanishes, i.e. ${\bf E}_a$ = ${\bf H}_{ab}$ ${\bf u}^b$ = 0. This is the frozen-in condition of the magnetic field [6]. Alternatively, one may write\\
\begin{equation}
{\bf H}^{ab} = {\bf {\eta}}^{abcd}_{{B_c}{u_d}} \hspace{1.0cm} {\Leftrightarrow} \hspace{1.0cm} {\bf H}^{*ab} = 2{{\bf u}^{[a} {\bf B}^{b]}},
\end{equation}
\noindent where the magnetic induction vector ${\bf B}_a$ satisfies the constitutive relation ${\bf B}_a$ = $\mu$ ${\bf H}_a$. The permeability $\mu$ is assumed to be constant. ${\bf H}_a$ denotes the magnetic field. It is evident from (4) that ${\bf H}^{ab}$ ${\bf H}^*_{ab}$  = 0. This is well-known magnetohydrodynamic (MHD) condition. It is apparent from (4) that ${\bf H}^{*ab}$  is the skew-symmetric product of a pair of linearly independent vectors ( ${\bf u}^a, {\bf B}^a$ ). The vectors ( ${\bf u}^a$, ${\bf B}^a$ ) span a 2-dimensional vector space which is usually referred to as a blade of the bivector  ${\bf H}^{*ab}$ . Because of MHD condition ${\bf H}^{*ab}$  is a simple bivector of second rank. Since ${\bf u}^a$ is timelike and ${\bf B}^a$ is spacelike, the simple 
bivector ${\bf H}^{*ab}$  is timelike. Such a simple timelike bivector ${\bf H}^{*ab}$  has been called a "magnetic blade" [7]. Since the contraction of (1) with ${\bf H}_{ca}$ yields the necessary and sufficient frobenius condition [8] for ${\bf H}^{*ab}$   to be 2-surface forming, there exists a family of timelike 2-surfaces spanned by the fluid flow lines and the magnetic field lines [9]. Thus the magnetic blades mesh together to form a well-defined family of timelike 2-surfaces which we denote by{\bf M}. From now on, we shall call each {\bf M} a "magnetic world sheet". As is known that a string is a spacelike curve and it traces out a timelike world sheet when it is in motion [10]. Because of the frozen-in magnetic field, the fluid particles once locked to a magnetic field line will remain so throughout the motion and hence each "magnetic world sheet" M can be thought of as a string. Following Carter [11], we introduce a unit time-like bivector associated with {\bf M} as \\
\begin{center}
\begin{equation}  
{\epsilon}^{ab}  = 2{{\bf u}^{[a}} {\bf n}^{b]} 
\end{equation}
\end{center}
\noindent with the properties that 
\\
\begin{equation}  
{\epsilon}^{ab} {{\epsilon}_{ab}} = -2, \hspace{1.0cm} {{\bf u}^a} {{\bf n}_a} = 0 ,
\end{equation}
\noindent where ${\bf n}^a$ denotes the unit spacelike vector field whose integral curves 
represent the magnetic field lines. 
A straight forward but simple calculation gives 

\begin{equation}  
{\epsilon}^*_{ca} {\epsilon}^{*ab}_{;b}  =  {{\bf K}_c} ,
\end{equation}
		 						
\noindent where  ${{\epsilon}^*_{ca}}$ is the dual of ${{\epsilon}_{ca}}$ and ${{\bf K}_c}$ is given by\\
	 		
\begin{equation}  
{\bf K}_c = {{\bf n}^{\oplus}_c} - {{\bf u}^{\bullet}_c} + {{\bf n}_c} {{\bf u}^{\bullet}_a} {{\bf n}^a} + {{\bf u}_c}{{\bf n}^{\oplus}_a}{{\bf u}^a}
\end{equation}

\noindent Here an overhead circumflex $({\oplus})$ and dot (${\bullet}$) are, respectively, used to 
indicate the directional derivatives along the unit vector field ${\bf n}^a$ and the 
fluid 4-velocity ${\bf u}^a$. We call ${\bf K}_c$ the extrinsic curvature vector of the "magnetic world sheet" {\bf M} in the terminology of Carter and Langlois 
[12]. Setting ${\bf H}^{ab}$ = - ${\bf H}{\epsilon}^{ab}$  and substituting it into (2), we find that\\ 
\begin{equation}  
{{\bf H}_{,b}}{{\epsilon}^{*ab}} + {{\bf H}{{\epsilon}^{*ab}_{;b}}} = -{{\bf J}^a} ,
\end{equation}
\noindent Contracting (9) with ${\epsilon}^*_{ca}$  and using (7), we get\\
\begin{equation}  
{\bf HK}_c +{\bf Z}_c = {\bf H}_{,b} {\hat{\gamma}}^b_{c,} ,
\end{equation}
\noindent where 
\begin{equation}  
{\bf Z}_c = {\epsilon}^*_{ca} {({\jmath}^a_{(n)} + {\jmath}^a_{(s)})} , \hspace{1.0cm} {\hat{\gamma}^b_c} = {\delta}^b_c + {\bf u}^b {\bf u}_c -{\bf n}^b {{\bf n}_c}
\end{equation}
 
\noindent It is interesting to note that (10) bears a striking resemblance with that of the equation of motion for a global cosmic string in an axion field 
background [12]. ${\bf Z}_c$ represents the Lorentz force per unit length of a 
magnetic tube acting on each individual magnetic tube. This force isinterpretale as a magnetic analogue of Joukowski force (or Magnus force) that has been recognized in the context of vortex tube [12]. The 
magnetic field intensity ${\bf H}$ acts as a tension in the string interpretation of 
the magnetic field lines.\\
\section {\bf Tie-up with the Rotation of Magnetic Field Lines and Superconducting Currents}
\noindent In this section we confine our attention to seek a relation between supercurrent and rotation of thin magnetic tube trapped in the type II superconducting regions of protons under the flux freezing condition. Due to lack of space we omit discussions related to supercurrents which can be found in Ahmedov [13]. Substituting the second relation of (4) into (2) and inverting the resulting equation with the help of (1), we find that\\
\begin{equation}  
{\bf {\eta}}^{abcd}_{{H}_b {H}_{c;d}} = 2{\eta}^{abcd}_{{H}_b{u}_c} {\sigma}_{de} {\bf H}^e -{({\bf H}_b {\jmath}^b_{(s)})} {\bf u}^a ,
\end{equation}
\noindent where ${\sigma}_{de}$de denotes the shear of the fluid. Setting ${\bf H}_a$ = ${\bf Hn}_a$ and applying Greenberg's theory of spacelike congruence [14] for the congruence of magnetic field lines, we arrive at 
\begin{equation}  
{\bf {\eta}}^{abcd}_{{H}_b {H}_{c;d}} = 2 {H}^2{\bf R}^a + 2{\eta}^{abcd}_{{H}_b{u}_c} {\sigma}_{de} {\bf H}^e , 
\end{equation}
\noindent where ${\bf R}^a$ denotes the rotation of the congruence of magnetic field lines that generates a magnetic tube. It follows from (12) and (13) that \begin{equation}  
{\bf R}^a = - \left[{\frac{{{\bf H}_b}{{\jmath}^b_{(s)}}} {2 {{\bf H}^2}}}\right] {\bf u}^a 
\end{equation}
\noindent which leads to 
\\
\begin{equation}  
{\bf R}^2 = - \left[\frac{{\bf H}_b{\jmath}^b_{(s)}} {2{{\bf H}^2}}\right]^2, \hspace{1.0cm} {{\bf R}^2} = -{\bf R}^a{{\bf R}_a}   
\end{equation}
\noindent It is evident from (15) that the superconducting current causes localrotation of a magnetic tube. This implies that the magnetic flux tubes carryin quantized flux are locally rotating thin tubes in type II superconductor. This local rotation remains operative as long as the induced magnetic field
 strength (due to the normal conduction currents) is above the lower critical field but below the upper critical field (i.e. ${{\bf H}_{c1}}$  $\leq$  ${\bf H}$  $\leq$   ${\bf H}_{c2}$ )[15]. \\
\begin{center}
{\bf Acknowledgement}      
\end{center} 
\noindent The authors would like to thanks Inter-University Centre for Astronomy and Astrophysics, Pune for providing facility where this work was carried out We acknowledge the co-operation of S.C.College, Ballia and M.B.Govt.  P.G. College, Haldwani, authority. \\

\end{document}